\documentclass[english,twocolumn,showpacs,preprintnumbers,groupedaddress,amsmath,amssymb,aps,pra]{revtex4}
\usepackage[T1]{fontenc}
\usepackage[latin1]{inputenc}
\usepackage{textcomp}
\usepackage{graphicx}
\usepackage{amssymb}
\usepackage{esint}

\makeatletter

\@ifundefined{textcolor}{}
{%
 \definecolor{BLACK}{gray}{0}
 \definecolor{WHITE}{gray}{1}
 \definecolor{RED}{rgb}{1,0,0}
 \definecolor{GREEN}{rgb}{0,1,0}
 \definecolor{BLUE}{rgb}{0,0,1}
 \definecolor{CYAN}{cmyk}{1,0,0,0}
 \definecolor{MAGENTA}{cmyk}{0,1,0,0}
 \definecolor{YELLOW}{cmyk}{0,0,1,0}
 }

\makeatletter\usepackage{bm}\makeatother\usepackage{babel}

\makeatother

\usepackage{babel}

\begin{document}

\title{Optical Precursors with Self-induced Transparency}

\author{Bruno Macke}

\author{Bernard S\'{e}gard}

\email{bernard.segard@univ-lille1.fr}

\affiliation{Laboratoire de Physique des Lasers, Atomes et Molécules (PhLAM),
Centre d'Etudes et de Recherches Lasers et Applications, CNRS et Université
Lille 1, 59655 Villeneuve d'Ascq, France}

\date{\today}
\begin{abstract}
Optical Sommerfeld-Brillouin precursors significantly ahead of a main
field of comparable amplitude have been recently observed in an opaque
medium with an electromagnetically induced transparency window {[}Wei
\emph{et al}., Phys. Rev. Lett. \textbf{103}, 093602 (2009){]}. We
theoretically analyze in this paper the somewhat similar results obtained
when the transparency is induced by the propagating field itself and
we establish an approximate analytic expression of the time-delay
of the main-field arrival, which fits fairly well the result obtained
by numerically solving the Maxwell-Bloch equations.
\end{abstract}


\maketitle
More than one century ago, Sommerfeld examined the apparent inconsistency
between the existence of superluminal group velocities and the theory
of relativity. Considering an incident field switched on at time $t=0$
(step pulse), he showed that, no matter the value of the group velocity,
no field can be transmitted by a linear dispersive medium before the
instant $t=L/c$ where $L$ is the medium thickness and $c$ the velocity
of light in vacuum \cite{som07}. Subsequently he and Brillouin studied
the fast oscillatory transients appearing at $t\geq L/c$ in the particular
case of a single-resonance Lorentz medium \cite{som14,bri14}. They
named them forerunners insofar as, in proper conditions, they can
distinctly precede the establishment of the steady-state field (the
main field). Renamed optical precursors, forerunners have entered
classical textbooks \cite{stra41,jack75} and continue to raise a
considerable interest. The theoretical results of Sommerfeld and Brillouin
have been improved, even rectified (the amplitude of the precursors
was in particular strongly underestimated in their work), and different
models of linear dispersive media have been considered. See \cite{oug07}
for a recent review.

Despite the abundant literature on precursors, there are very few
papers reporting direct observation of precursors distinguishable
from the main field. The difficulty of such an observation has been
soundly discussed by Aavikssoo \emph{et al}. \cite{aa88} who achieved
in 1991 an experiment involving single-sided exponential pulses (instead
of step-pulses) and exploiting the dispersion originating from a narrow
exciton line in AsGa \cite{aa91}. For proper detuning of the optical
carrier frequency $\omega_{c}$ from the resonance frequency $\omega_{0}$,
optical precursors appear as a small spike superimposed on the main
pulse. See also \cite{jeon06,du08}. The observation of precursors
significantly ahead of a main field of comparable amplitude obviously
requires the use of long enough square pulses and of a medium fairly
transparent at the optical carrier frequency, the corresponding group
delay being long compared to the duration of the precursors. As discussed
in \cite{jeon09,bm09}, the latter conditions are met in an opaque
medium with a narrow transparency window (slow light medium). Such
an experiment has been recently achieved by Wei \emph{et al}. \cite{wei09}
in an opaque cloud of cold atoms with an electromagnetically induced
transparency (EIT) window. Note that, in this experiment (as in all
the studies of precursors), the propagating field linearly interacts
with the medium.

\begin{figure}[h]
\begin{centering}
\includegraphics[width=80mm]{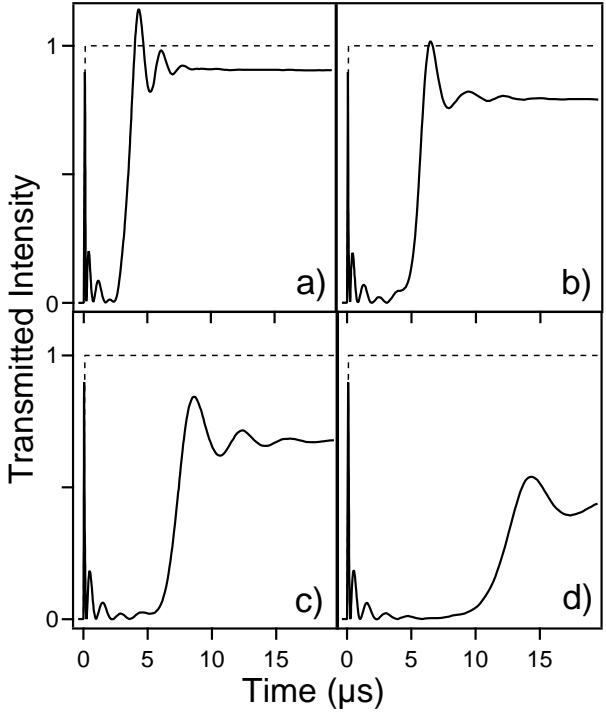} 
\par\end{centering}

\caption{Observed step response of a resonant absorbing medium. Parameters
: $\alpha L\approx200$ , $T_{1}\approx T_{2}\approx10\:\mu \mathrm{s}$ ($T_{2}/\alpha L\approx50\: \mathrm{ns}$), $T_{2}^{*}\approx1.3\:\mu \mathrm{s}$ , $\omega_{c}=5.4\times10^{11}\mathrm{s}^{-1}$
($\omega_{c}^{-1}\approx1.8\: \mathrm{ps}$ ), $\tau_{r}=12\: \mathrm{ns}$ ; $I_{0}\approx$
(a) 2100,(b) 960 (c) 620 (d) 350. In each case, the intensity is normalized to that of the step transmitted in the absence of gas (dashed line).\label{fig:FigExperiments}}

\end{figure}

For comparison, we will examine here the nonlinear situation where
\emph{the medium transparency is induced by the propagating field
itself} \cite{mcc69,crisp72}. Figure \ref{fig:FigExperiments}
shows the result of an experiment achieved in such conditions \cite{bs90}.
The medium is a gas of $\mathrm{HC^{15}N}$ at low pressure contained
in a 182m-long oversized waveguide and the incident wave is on resonance
with the molecular rotational line $J=0,\, M=0\rightarrow J=1,\, M=0$
(wavelength $\lambda_{c}\approx3.5$ mm). The gas behaves as a 2-level
medium \cite{all87} characterized by $T_{1}$ ($T_{2}$ ) the relaxation
time for the population difference (the polarization), $T_{2}^{*}$
the Doppler time and $\alpha$ the resonant absorption coefficient
at low intensity (extrapolated from the Lorentzian wings of the line).
See \cite{bs89} for details. The incident wave is characterized by
$I_{0}$ its intensity normalized to the saturation intensity and
$\tau_{r}$ its rise time. The observed step responses clearly have
some similarities with those obtained in the EIT experiment \cite{wei09},
with a short transient preceding the establishment of a steady state
regime (main field). The quasi Rabi oscillations \cite{crisp72} accompanying
the latter are obviously absent in the EIT experiments but oscillations
having a linear origin (postcursors) can also be observed in this
case \cite{bm09}.

To analyze the previous results, we provisionally neglect the Doppler
broadening and assimilate the guided wave to a plane wave propagating
in the $z$ direction ($0<z<L$), with an electric field polarized
in the $x$ direction. As long as $\tau_{r},\, T_{1},\, T_{2}\gg1/\omega_{c}$
and $\alpha\ll\omega_{c}/c$, the slowly varying envelope approximation
(SVEA) \cite{all87} holds \cite{re1} and we write the $E_{x}$ component
of the electric field as :
\begin{equation}
E_{x}(z,t)=\mathrm{Re}\left[\mathrm{e}^{i\omega_{c}t}\widetilde{E}(z,t)\right]\label{eq1}
\end{equation}
where, as in all the following, $t$ is a \emph{local time} (real time minus
$z/c$), and $\widetilde{E}(z,t)$ is the slowly varying field envelope.
Denoting $\mu$ the dipole matrix element for the transition (chosen
real), $R(z,t)=\mu\widetilde{E}(z,t)/\hbar$ the Rabi frequency, $n(z,t)$
the population difference per volume unit ($n_{0}$ its value at equilibrium)
and $\widetilde{P}(z,t)$ the envelope of the electric polarization
induced in the medium, it is convenient to introduce the dimensionless
quantities $D=n/n_{0}$, $P=\frac{i\widetilde{P}}{n_{0}\mu}\sqrt{\frac{T_{1}}{T_{2}}}$
and $E=\mu\widetilde{E}\sqrt{T_{1}T_{2}}/\hbar=R\sqrt{T_{1}T_{2}}$,
all real in the resonant case. $I=E^{2}$ is the intensity normalized
to the saturation intensity. The Maxwell-Bloch (MB) equations governing
the system evolution take then the simple form
\begin{equation}
\frac{\partial E}{\partial z}=-\frac{\alpha}{2}P\label{eq2}
\end{equation}
\begin{equation}
T_{2}\frac{\partial P}{\partial t}=DE-P\label{eq3}
\end{equation}
\begin{equation}
T_{1}\frac{\partial D}{\partial t}=-PE+(1-D)\label{eq4}
\end{equation}

We assume that the rise time $\tau_{r}$ of the incident intensity,
while long compared to $1/\omega_{c}$ (as above mentioned), is short
with respect to all the other characteristic times of the system ($1/R$,
$T_{1}$, $T_{2}$ and $T_{2}/\alpha L$). The response $E(L,t)$
of the medium (with a time resolution equal to $\tau_{r}$) is then
obtained by solving the MB equations with $P(z,0)=0$, $D(z,0)=1$
and $E(0,t)=E_{0}\Theta(t)=\sqrt{I_{0}}\Theta(t)$ where $\Theta(t)$
is the unit step function. This problem has been examined by Crisp
\cite{crisp72} when the relaxation effects are negligible, a condition
obviously not met in the experiments.

The long term behavior of the step response ($t\gg T_{1},T_{2}$ )
is obtained by solving the MB equations in steady state. Combining
Eqs.\ref{eq3} and \ref{eq4}, we find $P=E/(1+E^{2})$ and, putting
this result in Eq.\ref{eq2}, we easily retrieve the transmission
equation \cite{sel67,hil83,bm08}
\begin{equation}
I(\infty)+\ln I(\infty)=I_{0}+\ln I_{0}-\alpha L\label{eq5}
\end{equation}
where $I(t)$ is a short hand notation of the transmitted intensity
$I(L,t)$. The medium being optically thick in the linear regime ($\alpha L\gg1$),
the absorption is fully saturated ($I(\infty)/I_{0}\approx1$) only
when the incident (normalized) intensity is extremely large ($I_{0}\gg\alpha L$).
In fact the transmitted field (main field) will be significant (partial
transparency) as soon as $I_{0}-\alpha L=\mathrm{O}(\alpha L)$. The transmission
equation takes then the approximate form $I(\infty)/I_{0}\approx1-\alpha L/I_{0}$
and a transmission $I(\infty)/I_{0}>1/3$ is obtained for $I_{0}>3\alpha L/2$. 

Consider now the short term behavior of the step response. By combining
the integral form of Eqs.\ref{eq3} and \ref{eq4} and taking into
account that $D(z,t)\leq1$, one can establish the inequality \cite{crisp70}
\begin{equation}
1-D(z,t)<\left|\int_{0}^{t}R(z,t')dt'\right|^{2}<R_{0}^{2}t^{2}\label{eq6}
\end{equation}
where $R_{0}$ is the Rabi frequency associated with the incident
step ( $R_{0}^{2}=\frac{I_{0}}{T_{1}T_{2}}$). When $R_{0}^{2}t^{2}\ll1$,
$D(z,t)\approx1$ and the MB equations are reduced to the couple of
linear equations $\partial E/\partial z=-\alpha P/2$ and $T_{2}\partial P/\partial t=E-P$.
So, at least in this time domain and though $I_{0}\gg1$, the medium
behaves as a linear system (small pulse-area approximation \cite{crisp70}).
Its response $E(L,t)$ is easily retrieved from the previous couple
of equations and can be written as \cite{lau78,bs87}
\begin{equation}
E(L,t)=E_{0}\Theta(t)\left(1-\alpha L\intop_{0}^{t/T_{2}}\frac{\mathrm{J}_{1}\left(\sqrt{2\alpha Lu}\right)}{\sqrt{2\alpha Lu}}\mathrm{e}^{-u}du\right)\label{eq7}
\end{equation}
When $\alpha L\gg1$, the integral can be transformed to obtain
\begin{equation}
E(L,t) \approx E_{0}\Theta(t)\mathrm{e}^{-t/T_{2}}\mathrm{J}_{0}\left(\sqrt{2\alpha Lt/T_{2}}\right)\label{eq8}
\end{equation}
 For $x>1$, $\mathrm{J}_{0}(x)\approx\sqrt{\frac{2}{\pi x}}\cos\left(x-\frac{\pi}{4}\right)$
and $E(L,t>\frac{T_{2}}{2\alpha L})\approx E_{+}(t)+E_{-}(t)$ where 
\begin{equation}
E_{\pm}(t)=\frac{E_{0}}{\sqrt{2\pi}}\mathrm{e}^{-t/T_{2}}\frac{\exp\left[\pm i\left(\sqrt{2\alpha Lt/T_{2}}-\pi/4\right)\right]}{\left(2\alpha Lt/T_{2}\right)^{1/4}}\label{eq9}
\end{equation}
So the optical field is made of two components of equal amplitude
and instantaneous frequency $\omega_{c}\pm\sqrt{\frac{\alpha L}{2tT_{2}}}$,
which are nothing but that the Sommerfeld ($E_{+}$) and Brillouin
($E_{-}$) precursors as determined by the saddle point method of
integration \cite{lef09,bm09,re2}. The linear character of the short-term
response (and thus its analysis in terms of precursors) is well supported
by the experiments. As shown Fig.\ref{fig:FigExperiments}, the shape
of the corresponding transient is roughly independent of the incident
intensity. By numerically solving the MB equations, we find that the
condition $R_{0}^{2}t^{2}\ll1$ is much too severe and that the linear
approximation satisfactorily holds up to $t=2\pi/R_{0}$, the Rabi
period of the incident field. It even holds later in the experiments
because the transversal inhomogeneity of the field partially washes
out the (nonlinear) quasi Rabi oscillations while it does not affect
the linear response (the precursors). 

\begin{figure}[h]
\begin{centering}
\includegraphics[width=80mm]{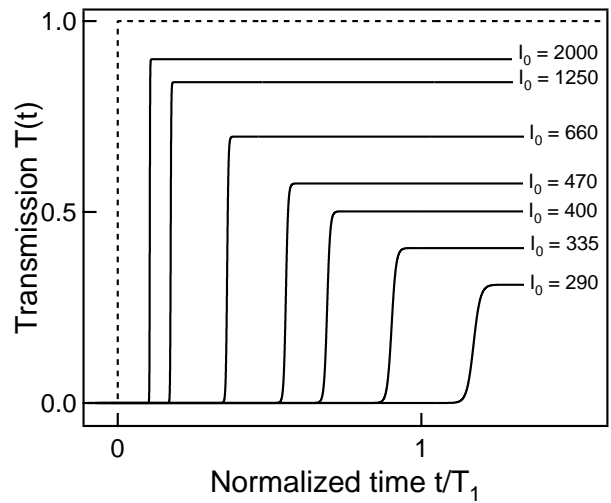}
\par\end{centering}
\caption{Step response obtained in the rate equations approximation (REA) as
a function of the normalized time $t/T_{1}$. Optical thickness $\alpha L=200$.
Each step response is labeled by the corresponding incident intensity
$I_{0}$. The step response obtained for $I_{0}\rightarrow\infty$
is given for reference (dashed line)\label{fig:StepResponseREA}.}
\end{figure}

In the EIT experiments, the probe field linearly interacts with the
medium at every time and the arrival of the main field is determined
by the (slow) group velocity \cite{bm09}. In the present case, this
arrival is fixed by fully nonlinear phenomena, the study of which
requires the resolution of the complete MB equations. We first examine
the solution obtained in the rate equations approximation (REA) \cite{all87}.
The equations to solve are then reduced to $\partial I/\partial z=-\alpha ID$
and $T_{1}\partial D/\partial t=-D((1+I)+1$ \cite{sel67,bm08} with
$D(z,0)=1$ and $I(0,t)=I_{0}\Theta(t)$. Eliminating $D$ and integrating
in $z$, we get \cite{bm08}
\begin{equation}
T_{1}d\left(\ln I\right)/dt=\ln I_{0}+I_{0}-\alpha L-\ln I-I
\label{eq10}\end{equation}
with $I(0)=I_{0}\exp\left(-\alpha L\right)$. The transmitted intensity
$I(t)$ is finally given by the implicit equation
\begin{equation}
\frac{t}{T_{1}}=\intop_{I_{0}\exp\left(-\alpha L\right)}^{I(t)}\frac{dI'}{I'\left(\ln I_{0}+I_{0}-\alpha L-\ln I'-I'\right)}
\label{eq11}\end{equation}
The transmission $T(t)=I(t)/I_{0}$ monotonously increases from $\exp\left(-\alpha L\right)$
to $I(\infty)/I_{0}$, where $I(\infty)$ is given by Eq.\ref{eq5}.
In the conditions considered here {[}$\alpha L\gg1$, $I_{0}-\alpha L=\mathrm{O}(\alpha L)${]},
$T(0)\approx0$, $T(\infty)\approx1-\alpha L/I_{0}$ and the transition
between these two values is very steep (Fig.\ref{fig:StepResponseREA}).
The time-delay of the arrival of the main intensity is conveniently
defined as the time $\tau_{d}$ such that $I(\tau_{d})=I(\infty)/2$.
It is given by Eq.\ref{eq11} by taking $I(\infty)/2$ as upper limit
of integration. When $I_{0}\gg\alpha L$ (full saturation limit),
Eq.\ref{eq11} can be explicitly integrated to give $T(t)\approx\left[1+\exp\left(\alpha L-I_{0}t/T_{1}\right)\right]^{-1}$
in agreement with the result given in \cite{kry70}. The 10-90\% rise
time $\Delta t$ of the intensity and the time-delay $\tau_{d}$ then
read as $\Delta t\approx4\ln3\left(T_{1}/I_{0}\right)$ and $\tau_{d}\approx\alpha L\left(T_{1}/I_{0}\right)\ll T_{1}$.
When the saturation is only partial, the time-delay $\tau_{d}$ as
a function of $1/I_{0}$ increases much faster than $\alpha LT_{1}/I_{0}$
and values of the order of $T_{1}$ can be attained while keeping
a significant transmission (Fig.\ref{fig:StepResponseREA}).

\begin{figure}[h]
\begin{centering}
\includegraphics[width=80mm]{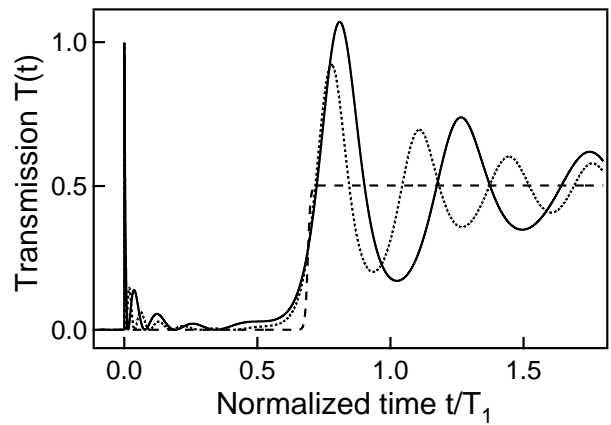}
\par\end{centering}
\caption{Comparison of the step response obtained by the REA (dashed line)
to those obtained by numerically solving the Maxwell-Bloch (MB) equations
with $T_{2}/T_{1}=1$ (full line) and with $T_{2}/T_{1}=1/2$ (dotted
line). Other parameters: $\alpha L=200$ and $I_{0}=400$, leading
to $T(\infty)\approx1/2$. Note that the pseudo-period of the oscillations
superimposed to the steady state in the MB solutions are nearly equal
to the corresponding Rabi period $2\pi/R_{\infty}=2\pi\sqrt{T_{1}T_{2}/I(\infty)}$,
namely $0.44T_{1}$ ($0.31T_{1}$) for $T_{2}/T_{1}=1$ ($T_{2}/T_{1}=1/2$).\label{fig:REAvsMB}}
\end{figure}

The REA does not take into account the coherent effects. It eliminates
in particular the quasi Rabi oscillations accompanying the main field.
One may however expect that the signals obtained by this way are a
satisfactory approximation of the exact signals, the oscillatory parts
of which would have been filtered out. To check this idea, we have
compared, $\alpha L$ and $I_{0}$ being fixed, the step response
obtained by using the REA (independent of $T_{2}$) to those obtained
by numerically solving the MB equations for two different values of
$T_{2}/T_{1}$ (Fig.\ref{fig:REAvsMB}). 
The three step responses are obviously different but the time-delays $\tau_{d}$ (as defined
before) are very close. Similar simulations made for different values
of the parameters show that this result is not accidental. It appears
that Eq.\ref{eq11} provides the exact time-delay with a precision
better than $10\%$ in all the cases of physical interest, that is
when the precursors are well developed before the arrival of the main
field and the latter has a significant amplitude.
\begin{figure}[h]
\begin{centering}
\includegraphics[width=80mm]{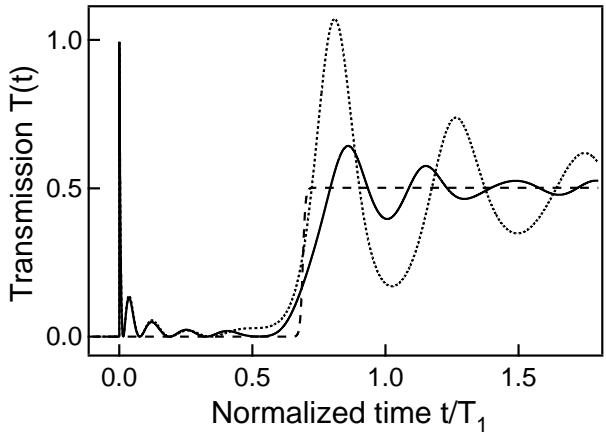}
\par\end{centering}
\caption{Numerical solution of the MB equations taking into account the transverse
distribution of the field (full line) for $\alpha L=200$, $I_{0}=400$,
and $T_{2}=T_{1}$. The REA (dashed line) and MB (dotted line) solutions
obtained with the plane wave model are given for reference.\label{fig:Transverse}}
\end{figure}

We will now examine the modifications brought to the step response
by some effects neglected in the previous theoretical analysis. The
most important one results from the transverse inhomogeneity of the
guided wave. Figure \ref{fig:Transverse} shows a typical step-response
obtained by using a MB numerical code extended to include a transverse
variation of the field \cite{pes91}. As expected, the linear part
of the response (precursors) is not changed (it is even slightly prolonged)
but the quasi Rabi oscillations (strongly depending on the field amplitude)
are dramatically affected. Their amplitude is considerably reduced
and their damping is accelerated, in agreement with the experimental
result (Fig.\ref{fig:FigExperiments}). However we remark that the
time-delay $\tau_{d}$ is not significantly larger than that obtained
in the plane-wave and rate-equations approximations. Similar calculations
including the Doppler broadening instead of the field inhomogeneity
in the plane-wave MB numerical code show that, even when $T_{2}^{*}=0.13T_{2}$
(parameters of Fig.\ref{fig:FigExperiments}), the Doppler effect
negligibly affects the precursors and slightly reduces the time-delay
$\tau_{d}$. This can be explained by observing that the right time
scale for the precursors and the nonlinear response is not $T_{2}$
but, respectively, $T_{2}/\alpha L\ll T_{2}^{*}$ and $1/R_{0}<T_{2}^{*}$.
Finally, the finite rise time of the incident step essentially affects
the most rapidly varying part of the step response, namely the transient
associated with the precursors and first the intensity $I_{1}$ of
its first peak. When $\alpha L\gg1$, $I_{1}$ only depends on $r=\alpha L\tau_{r}/T_{2}$
and attains the intensity $I_{0}$ of the incident wave when $r\ll1$.
This condition is approximately met in the experiment reported in
\cite{bs90} where $I_{1}\approx0.9I_{0}$ (Fig.\ref{fig:FigExperiments}).
Similar results could be obtained at optical wavelength by propagating a Gaussian beam in an ensemble of laser cooled 2-level atoms. We have then $T_{2}^{*}\gg T_{2}$  and the Doppler effect negligibly affects the precursors and the quasi Rabi oscillations. In other respects, $T_{2}$ (typically $50$ns) is about $200$ times shorter than in the microwave experiment. For a good observation of the precursors, the rise time of the incident step should also be $200$ times shorter, namely in the $50$ps range (attained with electro-optic modulators).

To summarize, we have shown that the experiments involving self-induced
transparency are a good alternative to the EIT experiments in order
to observe optical precursors well ahead of the main field, both having
intensities comparable to that of the step-modulated incident wave.
By using a plane-wave model and the rate equations approximation,
we have established an analytical expression for the time-delay of
the main-field arrival, which generalizes that previously obtained
in the infinite saturation limit, and we have shown that this expression
provides a good estimate of the real time-delay as long as precursors
and main field are well separated and of significant amplitude.


\begin{thebibliography}{29}
\bibitem{som07} A. Sommerfeld, Physikalische Zeitschrift \textbf{23},
841 (1907).

\bibitem[2]{som14} A. Sommerfeld, Ann. Phys. (Leipzig) \textbf{44},
177 (1914).  

\bibitem[3]{bri14} L. Brillouin, Ann. Phys. (Leipzig) \textbf{44},
204 (1914). 

\bibitem[4]{stra41} J.A. Stratton, \emph{Electromagnetic Theory}
(McGraw-Hill, New York 1941) 

\bibitem[5]{jack75} J.D. Jackson, \emph{Classical Electrodynamics},
2nd ed. (Wiley, New York 1975). 

\bibitem[6]{oug07} K.E. Oughstun, \emph{Electromagnetic and Optical
Pulse Propagation 1} (Springer, Berlin 2007), Ch.1. 

\bibitem[7]{aa88} J. Aaviksoo, J. Lippman and J. Kuhl, J. Opt. Soc.
Am. B \textbf{5}, 1631 (1988). 

\bibitem[8]{aa91} J. Aaviksoo, J. Kuhl, and K. Ploog, Phys. Rev.
A \textbf{44}, R5353 (1991).

\bibitem[9]{jeon06} H. Jeong, A. M. C. Dawes, and D.J. Gauthier,
Phys. Rev. Lett., \textbf{96}, 143901 (2006). 

\bibitem[10]{du08} S. Du, C. Belthangady, P. Kolchin, G.Y. Yin, and
S.E. Harris, Opt. Lett. \textbf{33}, 2149 (2008)

\bibitem[11]{jeon09} H. Jeong and S. Du, Phys. Rev. A \textbf{79},
011802(R) (2009).

\bibitem[12]{bm09} B. Macke and B. S\'{e}gard, Phys. Rev. A \textbf{80},
011803(R) (2009).

\bibitem[13]{wei09} Dong Wei, J.F. Chen, M.M.T. Loy, G.K.L. Wong,
and S. Du, Phys. Rev. Lett. \textbf{103}, 093602 (2009).

\bibitem[14]{mcc69} S.L. McCall and E.L. Hahn, Phys. Rev. \textbf{183},
457 (1969).

\bibitem[15]{crisp72} M.D. Crisp, Phys. Rev. A \textbf{5}, 1365 (1972).

\bibitem[16]{bs90} B. S\'{e}gard, B. Macke, J. Zemmouri, and W. Sergent,
Ann. Phys. (Paris), Colloque n\textdegree{}1, Supplément au n\textdegree{}2,
Vol.15 (1990), p.167.

\bibitem[17]{all87} L. Allen and J.H. Eberly, \emph{Optical resonance
and two-level atoms} (Dover, New York 1987).

\bibitem[18]{bs89} B. S\'{e}gard, B. Macke, L.A. Lugiato, F. Prati, and
M. Brambilla, Phys. Rev. A \textbf{39}, 703 (1989).

\bibitem[19]{re1} These conditions are met in all the experiments
having led to a direct observation of precursors.

\bibitem[20]{sel67} A. Selden, Br. J. Appl. Phys. \textbf{3}, 1935
(1967).

\bibitem[21]{hil83} L.W. Hillman, R.W. Boyd, J. Krasinski, and C.R.
Stroud, Opt. Commun. \textbf{45}, 416 (1983{]}. 

\bibitem[22]{bm08}B. Macke and B. S\'{e}gard, Phys. Rev. A \textbf{78},
013817 (2008).

\bibitem[23]{crisp70} M.D. Crisp, Phys. Rev. A \textbf{1}, 1604 (1970). 

\bibitem[24]{lau78} A. Laubereau and W. Kaiser, Rev. Mod. Phys. \textbf{50},
607 (1978).

\bibitem[25]{bs87} B. S\'{e}gard, J. Zemmouri, and B. Macke, Europhys.
Lett. \textbf{4}, 47 (1987).

\bibitem[26]{lef09} W.R. LeFew, S. Venakides, and D.J. Gauthier,
Phys. Rev. A \textbf{79}, 063842 (2009). 

\bibitem[27]{re2} This asymptotic method of integration is sometimes
opposed to the SVEA. In fact it can pertinently be used in the frame
of the latter \cite{bm09}.

\bibitem[28]{kry70} P.G. Kryukov and V.S. Letokhov, Sov. Phys. Uspekhi
\textbf{12}, 641 (1970). 

\bibitem[29]{pes91} E.M. Pessina, B. S\'{e}gard, and B. Macke, Optics
Commun. \textbf{81}, 397 (1991). 
\end{thebibliography}
\end{document}